\documentclass[12pt,preprint,longabstract]{aastex}
\newcommand{\text}[1]{\ensuremath{\mathrm{#1}}}
\usepackage{units}
\usepackage{epsfig}
\def\registered{{\ooalign{\hfil\raise .00ex\hbox{\scriptsize R}\hfil\crcr\mathhexbox20D}}}                         

\begin{document}
\title{Search for Astrophysical Neutrino Point Sources at Super-Kamiokande}

\author{
E. Thrane\altaffilmark{20},
K. Abe\altaffilmark{1},
Y. Hayato\altaffilmark{1,33},
T. Iida\altaffilmark{1},
M. Ikeda\altaffilmark{1},
J. Kameda\altaffilmark{1},
K. Kobayashi\altaffilmark{1},
Y. Koshio\altaffilmark{1,33},
M. Miura\altaffilmark{1},
S. Moriyama\altaffilmark{1,33}
% ...
}

\affil{The Super-Kamiokande Collaboration}

\altaffiltext {1}{Kamioka Observatory, Institute for Cosmic Ray Research, The University of Tokyo, Hida, Gifu 506-1205, Japan}
% ...
\altaffiltext{20}{Department of Physics and Astronomy, University of Minnesota, Minneapolis, MN, 55455, USA}
% ...
\altaffiltext{33}{Institute for the Physics and Mathematics of the Universe (IPMU), The University of Tokyo, Kashiwa, Chiba 277-8568, Japan}
% ...

\date{\today}

\begin{abstract}
  It has been hypothesized that large fluxes of neutrinos may be created in astrophysical ``cosmic accelerators.''
The primary background for a search for astrophysical neutrinos comes from atmospheric neutrinos, which do not exhibit the pointlike directional clustering that characterizes a distant astrophysical signal.
  We perform a search for neutrino point sources using the upward-going muon data from three phases of operation (SK-I, SK-II, and SK-III) spanning $\unit[2623]{days}$ of live time taken from April 1, 1996 to August 11, 2007.
  The search looks for signals from suspected galactic and extragalactic sources, transient sources, and unexpected sources.
  We find interesting signatures from two objects---RX J1713.7-3946 (97.5\% CL) and GRB 991004D (95.3\% CL)---but the signatures lack compelling statistical significance given trial factors.
  We set limits on the flux and fluence of neutrino point sources above energies of $\unit[1.6]{GeV}$.
\end{abstract}

\maketitle

\section{Introduction}
The most plausible explanation of ultra-high-energy (UHE) cosmic rays is the so-called cosmic accelerator model, which posits that protons are accelerated by electromagnetic energy associated with objects with extended magnetic fields.
It has been pointed out that cosmic accelerators, through pion production, may be capable of producing copious high-energy neutrinos (in excess of $\unit[1]{GeV}$) at fluxes detectable by current or planned experiments~\citep{WB}.

\begin{figure}
  \leavevmode
  \psfig{file=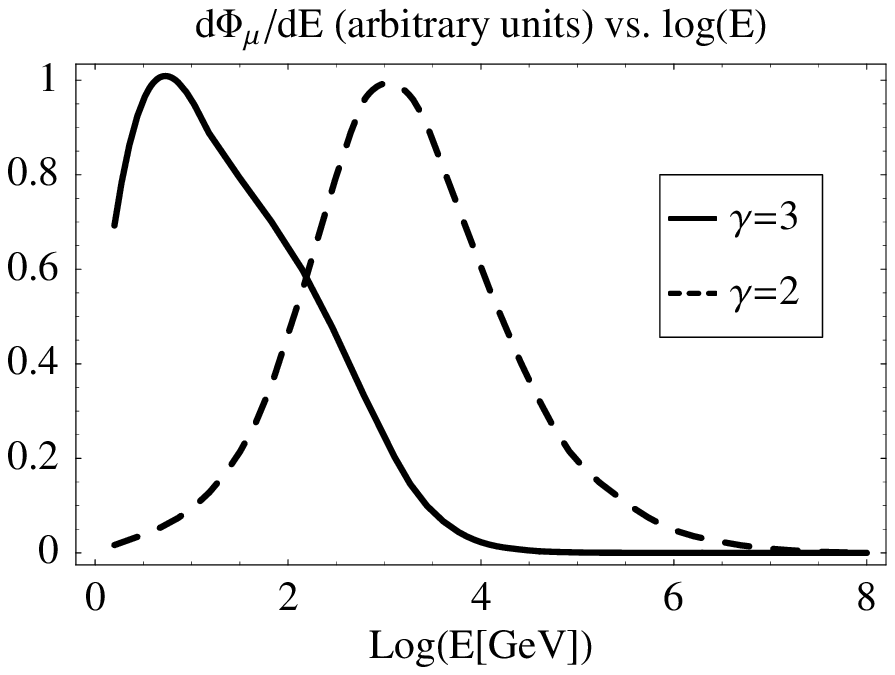,height=2.5in,width=3in}
  \caption{The spectra of neutrino-induced upmus for a source with spectral indices $\gamma=2$ (dashed) and $\gamma=3$ (solid) and a minimum upmu energy of $E_\mu^\text{min}=\unit[1.6]{GeV}$. The normalized spectra cross in the neighborhood of $\sim\unit[200]{GeV}$. \label{fig:upmuspec}}
\end{figure}

Many models invoke the physics of shocks, which predicts a power law spectrum that falls off like $d\Phi_\nu/dE\propto E^{-\gamma}$ with a spectral index of $\gamma\approx2$~\citep{LearnedMannheim}.
The cross section for neutrino-nucleon scattering, however, depends roughly linearly on energy, and the effective detector volume increases with energy since high-energy muons can travel further in rock than lower-energy ones.
For the most likely range of spectral indices ($\gamma = 2 \sim3$), the vast majority of astrophysical neutrino events at Super-Kamiokande fall into the category of upward-going muons described in Section~\ref{upmu}.
The spectrum of neutrino-induced muon events at Super-Kamiokande is shown in Figure~\ref{fig:upmuspec}.

\section{Upward-Going Muons at Super-Kamiokande}\label{upmu}
Super-Kamiokande is a water Cherenkov detector located in Mt. Ikeno in central Japan under 2700 meters water-equivalent rock overburden.
The cylindrically shaped detector contains $\unit[50]{ktons}$ of ultrapurified water.
It is divided by a steel structure into an inner detector (ID) equipped with 11146 $\unit[50]{cm}$ PMTs aimed inward and an optically separated outer detector (OD) instrumented with 1885 $\unit[20]{cm}$ PMTs aimed outward and equipped with wavelength-shifting plastic plates.
The primary function of the OD is to identify charged particles that enter or exit the ID most of which are cosmic rays.
Within the ID we define a $\unit[22.5]{kton}$ fiducial volume, within which detector response is expected to be uniform.
Further details regarding the Super-Kamiokande detector design, operation, calibrations, and data reduction can be found in References~\cite{Fukuda:2002uc,Ashie:2005ik}.

Neutrino events with total deposited energy above $\approx\unit[100]{MeV}$ are overwhelmingly due to atmospheric neutrinos from the decay of pions created by cosmic rays in the upper atmosphere.
To good approximation the Earth is transparent to neutrinos up to energies of order of $\unit[100]{TeV}$.
(The neutrino interactions observed by Super-Kamiokande represent a tiny fraction of the $\unit[1]{GeV}-\unit[100]{TeV}$ neutrinos passing through the Earth.)

\begin{figure}
  \leavevmode
  \psfig{file=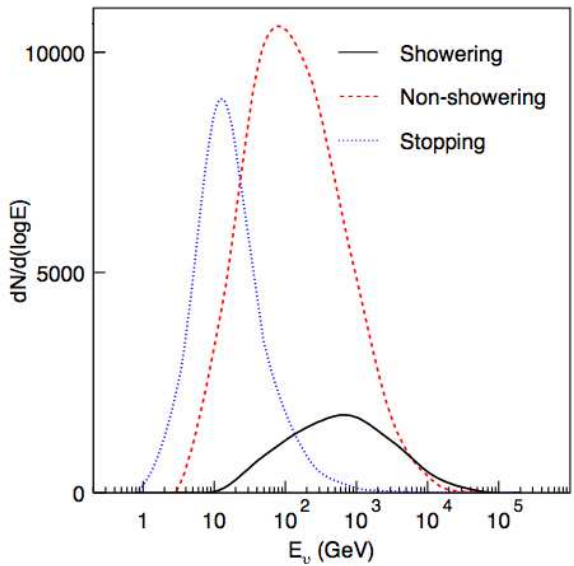,width=0.5\textwidth}
  \caption{The spectra of different event types at Super-Kamiokande obtained using $\unit[100]{years}$ of atmospheric neutrino MC (from~\citep{shantanu}). \label{fig:eventSpectra}}
\end{figure}

Super-Kamiokande classifies events according to their energy and/or topology (see Figure~\ref{fig:eventSpectra}).
Fully-contained (FC) neutrino events are those where interaction products are observed in the ID, with no significant correlated activity in the OD.
Partially-contained (PC) events are those where some interaction products exit the ID.
Upward-going muon (upmu) events occur when a penetrating particle traveling in the upward direction enters and either stops in or passes through the detector.
Upmu events are attributed to muons produced by neutrino interactions in the rock surrounding the detector.
A cut on reconstructed path length ensures that upmus have energies greater than $E^\text{min}_\mu=\unit[1.6]{GeV}$.
FC, PC, and upmu events are associated with successively higher energy samples of neutrino interactions, ranging from 200 MeV for the lowest energy FC events to above 1 TeV for the highest energy upmus.
We may thus think of each event type as a coarse energy bin so long as it is understood that the spectra for each event type overlap significantly.

Upmus are subdivided into stopping upmus, which stop inside the detector, and through-going, which pass straight through.
The through-going upmu sample is further divided into showering and non-showering upmus with characteristic neutrino parent energies of $\unit[800]{GeV}$ and $\unit[100]{GeV}$ respectively~\citep{shantanu}.
For point sources with spectral indices near $\gamma=2$, neutrinos are overwhelmingly likely to be observed in the through-going muon channel, and so we henceforth focus our attention on this dataset.
The upmu dataset includes a small contribution ($<1\%$) from nearly-horizontal cosmic rays, which constitute an additional source of background for an astrophysical search.
Other Super-Kamiokande astrophysical results making use of the upmu data include~\citep{shantanu,SKGRB,thrane080319B,molly,kristine}.
We address nearly-horizontal cosmic rays again in Section~\ref{likelihood}.

We utilize data from three stages of operation summarized in Table~\ref{tab:datasummary}.
The SK-II stage was characterized by diminished phototube coverage, (about half the value for SK-I and SK-III).
A comparison of solar-neutrino data from SK-I and SK-II has found no systematic differences in the two stages~\citep{Parker}.
The reduced photocoverage during SK-II caused the upmu angular resolution to worsen marginally: $1.46^\circ$ for SK-II in comparison with $1.05^\circ$ for SK-III.
(Angular resolution is defined such that 68\% of MC events have an angular separation between the true and reconstructed muon direction that is less than the resolution.)
We take this difference into account in Section~\ref{variables} in the construction of our search algorithm.
SK-I and SK-III are similar in that they share the same number of PMTs, but they differ in that SK-III has additional reflective material called Tyvek$^\registered$ in the OD to optically separate the caps from the barrel to aid in event reconstruction.

\begin{deluxetable}{l|c|r}
  \tabletypesize{\scriptsize}
  \tablecaption{A summary of the dataset for this analysis. (SK-III continued through 2008, but the data used here are limited to the dates indicated.) \label{tab:datasummary}}
  \tablewidth{0pt}
  \tablehead{ \colhead{phase} & \colhead{dates} & \colhead{events} }
  \startdata
    SK-I & Apr 1, 1996 - July 19, 2001 & 1,879 \\
    SK-II & Jan 17, 2003 - Oct 5, 2005 & 888 \\
    SK-III & Aug 4, 2006 - Aug 11, 2007 & 367 \\
  \enddata
\end{deluxetable}

\section{Likelihood Algorithm}\label{likelihood}
In this section we develop a likelihood algorithm to maximize the discovery potential.
We find it to be more sensitive by a factor of two than a previous search~\citep{shantanu} wherein we counted the number of showering events in a $4^\circ$ search cone, (see also~\citep{kristine}).
A similar algorithm has recently been presented as a means of searching for point sources with IceCube data~\citep{braun}.

To begin we construct an $8^\circ$ search cone around the test direction, which is big enough to include over 99\% of the point spread function (PSF).
We posit that any event, $i$, falling in the search cone has a probability $\alpha$ of being point-source signal, $S$, and a probability $(1-\alpha)$ of being atmospheric background, $B$, as in Equation~\ref{eq:prob}.
\begin{equation}\label{eq:prob}
  P_i=\alpha\,S(\theta_i...)+(1-\alpha)\,B(\theta_i...)
\end{equation}
$S(\theta_i...)$ is the probability density distribution for point-source neutrinos.
It depends, among other things, on the angular separation, $\theta$, from the search direction.
(The ellipses indicate that $S$ and $B$ depend on additional variables, which we shall enumerate in section~\ref{variables}.)

Smaller values of $\theta$ are more likely than large ones for the signal distribution.
$B(\theta_i...)$ is the probability density distribution for the background, which consists of atmospheric neutrinos plus a small contribution ($<1\%$) from nearly-horizontal cosmic ray muons that can masquerade as neutrino-induced upmus.
Larger values of $\theta$ are more likely than smaller ones for the background distribution since there is more phase space near the edge of the cone than at the center.

If there are $N$ events in the search cone, we can define a likelihood function, ${\cal L}$, as the product over events, $i$, of each $P(\alpha|\theta_i...)$ as in Equation~\ref{eq:likelihood}.
\begin{equation}\label{eq:likelihood}
  {\cal L}(\alpha)\equiv \prod_i^N P(\alpha|\theta_i...)
\end{equation}
Now we can interpret $\alpha$ as the fraction of events in the search cone due to signal as in Equation~\ref{eq:alpha}; when $\alpha=0$ there is no signal, when $\alpha=1$ the signal is maximal.
\begin{equation}\label{eq:alpha}
  \left\langle\alpha\right\rangle=N_S / (N_S + N_B)
\end{equation}
Here $N_S$ and $N_B$ are respectively the number of events in the cone due to signal and background.

As it stands, the likelihood function depends only on the properties of each event (such as $\theta$) and not on the excess/deficit of events in the search cone in comparison to the expected number of events given $\alpha$, which we denote $\bar{N}_{(\alpha)}$, and which is an additional measure of signal strength.
$\bar{N}(\alpha)$ is the {\it expected} number of events given $\alpha$ whereas $N$ is the number of {\it observed} events.

To add this information to the likelihood function, we define a generalized likelihood function, denoted ${\mathfrak L}$, which is the product of ${\cal L}$ with a Poisson probability distribution on $\alpha$, as in Equation~\ref{eq:generalized}.
\begin{equation}\label{eq:generalized}
  {\mathfrak L}_{(\alpha)}=
  P_\text{Poisson}\,{\cal L}_{(\alpha)}=
  e^{-\bar{N}(\alpha)} \,\frac{\bar{N}(\alpha)^N}{N!}\, \prod^N_{i=1} P_{(\alpha|\theta_i...)}
\end{equation}
Noting that $\bar{N}_{(\alpha)}=N_S+\bar{N}_B$, and recalling Equation~\ref{eq:alpha}, we see that $\bar{N}_{(\alpha)}=\bar{N}_B/(1-\alpha)$.
$\bar{N}_B$, the expected number of background events, can be determined by MC, and so we can eliminate all unknowns from Equation~\ref{eq:generalized} in order to obtain Equation~\ref{eq:general}.
\begin{equation}\label{eq:general}
  {\mathfrak L}_{(\alpha)}=\frac{e^{-\bar{N}_B/(1-\alpha)}}{N!} \left(\frac{\bar{N}_B}{1-\alpha}\right)^N {\cal L}_{(\alpha)}
\end{equation}

Having constructed the generalized likelihood function, we are now able to define a procedure for assessing signal strength.
Given a search direction from which we can define $\{\theta_i\}$, we vary $\alpha$ in order to find the best fit value, $\alpha_F$, which maximizes the generalized likelihood.
Finally, we define a likelihood ratio, $\Lambda$, as in Equation~\ref{eq:Lambda}, which serves as the ultimate indicator of signal strength.
\begin{equation}\label{eq:Lambda}
  \Lambda\equiv2\log\left[{\mathfrak L}(\alpha_F)/{\mathfrak L}(\alpha=0)\right]
\end{equation}
As $\Lambda$ gets larger, the probability of observing $\Lambda$ due to fluctuations in the atmospheric background, $P(>\Lambda)$, becomes smaller.
$P(>\Lambda)$ is determined numerically using atmospheric MC.

\section{Sensitivity}\label{sensitivity}
\begin{figure}
  \leavevmode
  \psfig{file=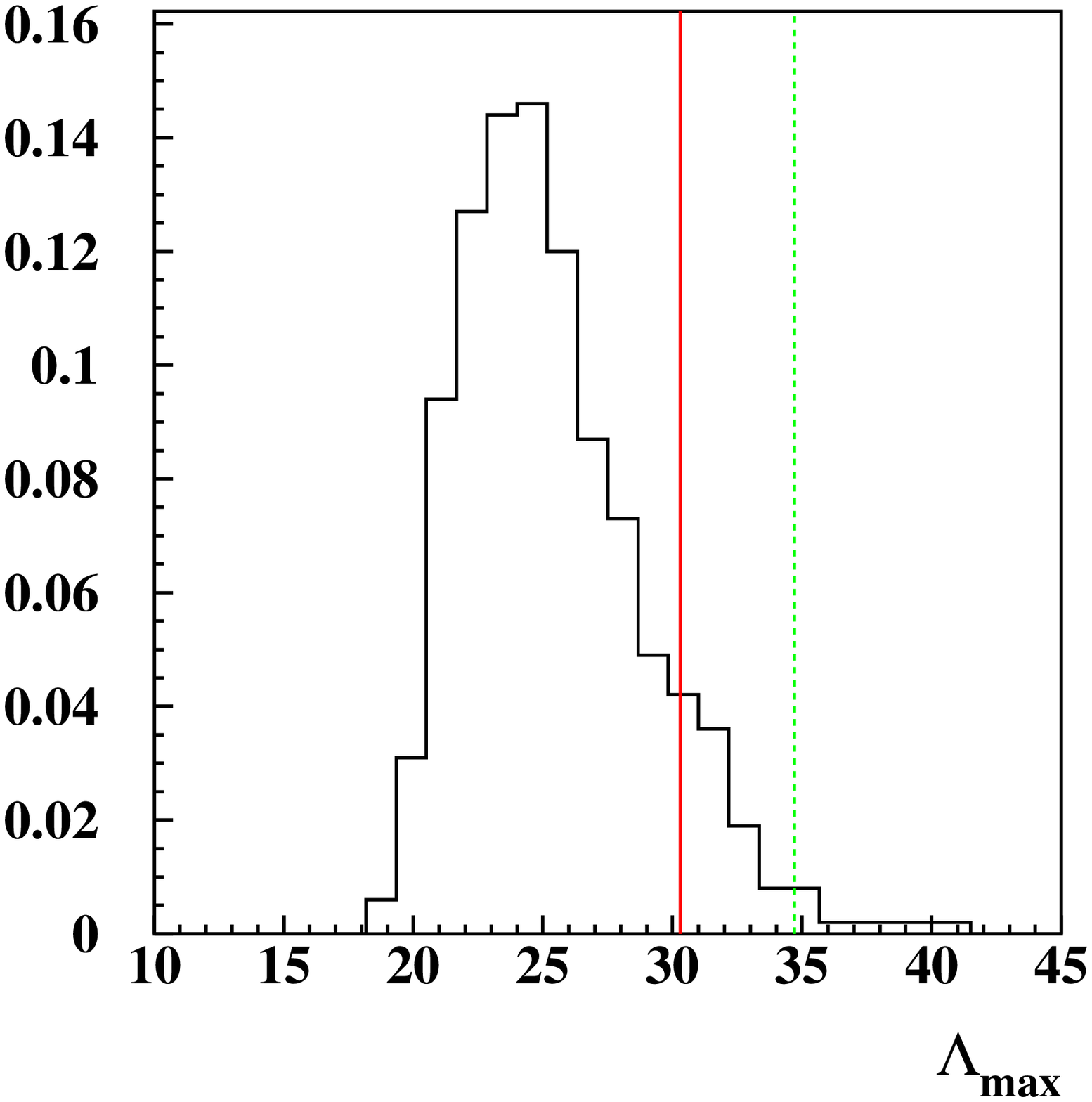,height=2.5in,width=3in}
  \caption{A normalized histogram of $\Lambda_\text{max}$. The thresholds for 90\%/99\% CL detection are respectively indicated with solid red and dashed green vertical lines.  Search directions are spaced at $0.5^\circ$ intervals. \label{fig:lambdamax}}
\end{figure}

In one test described in Section~\ref{searches}, we measure $\Lambda$ at regularly spaced $0.5^\circ$ intervals over the entire sky visible to Super-Kamiokande.
First we determine the largest signal at any point in the sky, $\Lambda_\text{max}$, and then we ascertain if it is large enough to constitute a statistically significant signal using the probability distribution, $P(\Lambda_\text{max})$, which is determined with MC.
Studying $P(\Lambda_\text{max})$ (see Figure~\ref{fig:lambdamax}) we find that the detection of a source at $\text{dec}=-15^\circ$ (in the middle of the range of declinations visible by Super-Kamiokande) at 90\% (99\%) CL with an efficiency of 50\% requires a signal of $\Lambda_\text{Max}=30.2$ $(34.7)$ respectively.
We make a distinction between the false alarm rate, which determines the threshold for $\Lambda_{90\%}$ $(\Lambda_{99\%})$, and the efficiency, which tells us the probability that a true signal will exceed the threshold.
These values of $\Lambda_\text{max}$ correspond to 10 (11) upmus out of 3134 total events.
Assuming a spectral index of $\gamma=2$, ten upmus corresponds to a neutrino flux of $\approx\unit[3\times10^{-7}]{cm^{-2}s^{-1}}$ above $\unit[1.6]{GeV}$.

The sensitivity calculated here describes the flux needed to observe a statistically significant signal by looking everywhere in the sky at regularly spaced $0.5^\circ$ intervals.
Such a search does not identify any a priori candidate sources, so we call it a ``tabula rasa search.''
It is important to note that the sensitivity of a search for a single a priori candidate is substantially better than the tabula rasa search because the latter samples the entire visible sky where the presence of fluctuations is sure to push the detection threshold higher.

A previous search for astrophysical neutrinos~\citep{shantanu} used the showering muon dataset to define search directions.
A $4^\circ$~cone was drawn around each search direction and the number of events inside the cone was compared to the number expected from atmospheric background.
The $4^\circ$~``hard cone search'' did not incorporate information about the angular separation of the events from the search direction and it did not make use of the non-showering dataset.
We find that the algorithm presented here requires half the signal of the hard cone test to obtain a signal at 90\% CL all other things equal.

\section{Upper Limit Calculation}
We also use the generalized likelihood function to calculate upper limits on the number of point-source neutrino-induced upmus as in Equation~\ref{eq:limits}.
\begin{equation}\label{eq:limits}
  \int_{\alpha_F-\delta}^{\alpha_{90}\equiv\alpha_F+\delta}{\mathfrak L}(\alpha) / \int_0^1{\mathfrak L}(\alpha) \equiv 90\% 
\end{equation}
The quantity $\alpha_{90}\equiv(\alpha+\delta)$ represents the upper limit at 90\% CL on the fraction of events in the search cone due to  signal.
(In the not infrequent event that $(\alpha_F-\delta)$ is unphysically less than zero, we modify Equation~\ref{eq:limits} so that the lower limit is 0.)
We thereby obtain an upper limit on the $E_\nu>\unit[1.6]{GeV}$ point-source neutrino-induced upmu flux as in Equation~\ref{eq:muflux}.
\begin{equation}\label{eq:muflux}
  \Phi_\mu=\frac{\alpha_{90}\,N}{A_\text{effective}\,t_\text{exposure}}
\end{equation}
Here $A_\text{effective}$ is the effective area of the detector and $t_\text{exposure}$ is the exposure time.
(Both $A_\text{effective}$ and $t_\text{exposure}$ depend implicitly on the search direction since they each vary with zenith angle, $z$.)
We defer discussion of the neutrino flux calculation until Section~\ref{nuflux}.

\section{Likelihood Variables}\label{variables}
Having introduced the formalism of the likelihood function, we are ready to give a full account of the variables on which it depends.
In addition to angular separation, $\theta$, the likelihood function depends on {\it event type}, denoted $n$, which can take on values of ``showering'' and ``non-showering.''
The atmospheric neutrino flux falls with a steep $d\Phi_\nu/dE\propto E^{-3.7}$ spectrum in the relevant energy range, whereas point-source neutrinos are thought to fall off like $d\Phi_\nu/dE\propto E^{-2}$.
Since showering muons come from a higher energy parent neutrino population than non-showering muons, showering muons provide stronger evidence of point-source neutrinos than non-showering muons.

\begin{figure}
  \leavevmode
  \psfig{file=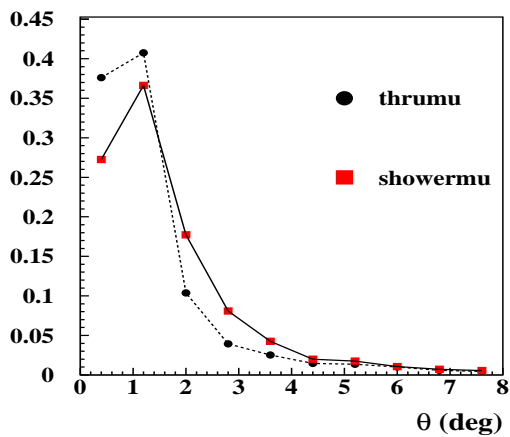,height=2.5in,width=3in}
  \caption{The normalized point spread function for showering (red squares, dashed line) and non-showering (black circles, solid line) muons includes contributions from scattering as well as detector resolution. Here we assume a spectral index of $\gamma=2$.  The vertical error bars are too small to see. \label{fig:psf}}
\end{figure}

The likelihood function also depends on $n$ since the PSF is different for showering and non-showering events.
(Showering events, which produce excess light through pair production, bremsstrahlung, and photonuclear effects, are on average slightly harder to reconstruct than non-showering events, which deposit energy primarily through ionization.)
An example of the PSF for showering and non-showering muons is depicted in Figure~\ref{fig:psf}.

The likelihood function also depends on {\it detector geometry}, denoted $m$, which can take on values of ``SK-I,'' ``SK-II,'' or ``SK-III.''
The PSF and showering algorithm efficiency both vary with $m$.
The likelihood additionally depends on each event's {\it zenith angle}, $z$, as well as the equatorial coordinates of the search direction: {\it right ascension}, $\text{ra}$, and {\it declination}, $\text{dec}$.
Dependence on $z$ arises from modest differences in the PSF at different zenith angles.
(The fitter performs best for slightly upward-going going muons and worst for straight upward-going muons due to detector geometry.)

Dependence on (ra,dec) enters through the number of expected background events in the cone, $\bar{N}_B$.
Atmospheric neutrinos have a well-studied zenith angle dependence due to the ``secant theta effect'' \citep{LearnedMannheim} from mesons interacting in the atmosphere plus additional effects from oscillations and the detector shape.
Each equatorial coordinate (ra,dec) is associated with a different locus of zenith angles and so the zenith angle distribution of atmospheric neutrinos manifests itself as a dependence on equatorial coordinates.
Thus, each measurement of signal strength is characterized by the following variables: 
\begin{equation}
  \prod_i^N\left\{\theta_i,n_i|m_i,z_i,\text{ra},\text{dec}\right\}
\end{equation}

In order to determine the signal and background distributions we use MC.
The background MC begins with atmospheric upmu vectors generated with NEUT~\citep{NEUT} and using the Honda model of atmospheric neutrino flux~\citep{Honda95}.
These upmu vectors become the input for a {\tt GEANT 3} based detector simulation.
The same MC has been described in previous Super-Kamiokande publications, e.g.~\cite{Ashie:2005ik}.
The output of this simulation is then processed by the upmu reduction as if it were real data.
We account for neutrino oscillations by applying a weighting factor proportional to the two-flavor survival probability, assuming maximal mixing and $\Delta m^2_{23}=\unit[2.5\times10^{-3}]{eV^2}$.
The procedure for accounting for contamination from nearly-horizontal cosmic rays using variations in the rock overburden is detailed elsewhere, (see, e.g., Reference~\cite{shantanu}.)

To generate signal MC we first assume that the source is characterized by a $d\Phi_\nu/dE\propto E^{-\gamma}$ power law with a spectral index of $\gamma=2$.
We obtain a spectrum for point source upmus by multiplying $d\Phi_\nu/dE$ by $P(E_\nu,\,E_\mu^\text{min})$---the probability that a neutrino with energy $E_\nu$ creates a muon with energy greater than $E_\mu^\text{min}=\unit[1.6]{GeV}$---and then integrating with respect to $E_\nu$ as in Equation~\ref{eq:nu2mu}.
(We explore this equation in greater detail in Section~\ref{nuflux} in the context of the neutrino flux calculation.)
Each upmu vector is assigned a scattering angle according to its energy before it is fed into the detector simulation, which is followed by the upmu reduction.

\section{Four Searches}\label{searches}
We conduct four searches for neutrino point sources.
The first search, mentioned in Section~\ref{sensitivity}, is the {\it tabula rasa} search.
For this search we make no assumptions about a priori suspected sources.
We measure signal strength $\Lambda$ at regular $0.5^\circ$ intervals over the entire sky visible to Super-Kamiokande and record the largest observed value, $\Lambda_\text{max}$, which we compare with MC to determine if a statistically significant signal is present.
This $0.5^\circ$ spacing is chosen to be smaller than the width of the point spread function (a little over a degree), but large enough to allow relatively fast computation.
The second search is one for {\it suspected candidates}.
In this search we consider sixteen objects identified in various publications as plausible bright sources of astrophysical neutrinos~\citep{Zhang,plerions,Huang,NEMOMQ,Link,Kistler,Bednarek,Gabici}.
Candidates include magnetars (young pulsars with strong magnetic fields and periods on the order or seconds); plerions (aka pulsar wind nebulae, nebulae with an embedded pulsar); supernova remnants (SNR, supernovae shockwaves colliding with surrounding gas); and microquasars (nearby black holes devouring a neighboring star).
A summary of the candidates tested is provided in Table~\ref{tab:candidates}.

In the third search, we test for a correlation between upmus and 27 {\it ultra-high-energy cosmic rays} observed by the Auger experiment, which may be linked to AGN~\citep{Auger,Auger2}.
We add the signal strength, $\Lambda$, from each direction to create a summed signal, $\Lambda_\Sigma$, the significance of which we evaluate with MC.

There are many ways to search for a correlation of neutrinos with AGN, but this method has several advantages.
There are more than 300 known AGN within $\unit[75]{Mpc}$ of Earth.
If we sum the signal from a great many cataloged AGN, we will effectively sample the entire sky, thereby diluting any signal.
It is not clear how to pick the most neutrino-bright AGN, or how many to include in a summed signal analysis.

The 27 UHE events observed by Auger, however, constitute a conveniently small dataset.
Also, it is plausible that the brightest cosmic ray objects are also the brightest in neutrinos.
Since the Auger experiment is a southern hemisphere experiment looking at downward-going cosmic rays, and Super-Kamiokande is a northern hemisphere experiment looking at upward-going muons, all 27 UHE events in the Auger dataset point to equatorial coordinates that are also visible to Super-Kamiokande.
UHE cosmic rays are deflected by electromagnetic fields, but for a reasonable choice of parameters this deflection is small, $\approx1^\circ$~\citep{lee}.
Finally, a test for a correlation of upmus with UHE cosmic rays is simple and transparent.

The fourth search is a search for upmus coincident with approximately 2200 GRBs in the BATSE~\citep{BATSE}, HETE~\citep{HETE}, and Swift~\citep{Swift} catalogs.
Following a previously established procedure~\citep{SKGRB}, we employ a $\pm\unit[1,000]{s}$ window centered on the beginning of photonic observations of each GRB.
This window size allows for any reasonable delay, positive or negative, between neutrino emission and photonic emission, while still providing a very small likelihood of random coincidences due to atmospheric neutrinos.
Since the upmu rate at Super-Kamiokande is only about $\unit[1.6]{day^{-1}}$, the timing cut alone provides a powerful filter to remove atmospheric background.
If there is a coincidence in time, we next determine if the event falls within the $8^\circ$ search cone.
If so, we evaluate the signal strength, $\Lambda$, of the coincident event.
The probability that the event is due to atmospheric background is given by the product of the probability of a coincidence in time, with the probability of a coincidence in space, with the probability of measuring a signal strength $\geq\Lambda$:
\begin{equation}
  P_\text{BG}=P_\text{time}\,P_\text{space}\,P(\geq\Lambda)
\end{equation}
As always, we evaluate $P(\geq\Lambda)$ with MC.

\section{Results}
\subsection{Tabula Rasa Search}
The tabula rasa search yielded a signal of $\Lambda_\text{max}=19.1$, which is well below the detection threshold.
In fact, this signal is very small given the distribution of $\Lambda_\text{max}$ in Figure~\ref{fig:lambdamax}, which prompted us to rerun the algorithm using ``boot-strapped'' values for local sidereal time.
``Boot-strapping'' in this context is the practice of assigning random values of local sidereal time to each event.
This has the effect of scrambling the equatorial coordinates associated with each event while preserving other properties such as $z$ and $n$.
The second boot-strap run yielded $\Lambda_\text{max}=24.9$, which, while still below the the detection threshold, is more typical, giving us confidence that the small value of $\Lambda_\text{max}=19.1$ is a reasonable fluctuation.
Since no signal was observed, we set upper limits on the flux of point-source neutrinos.

\begin{figure}
  \leavevmode
  \psfig{file=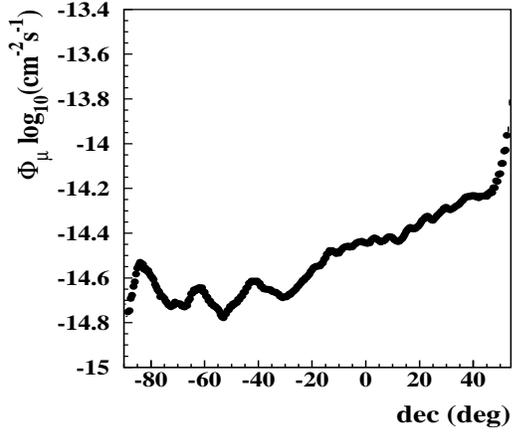,height=2.5in,width=3.0in}
  \caption{90\% CL limits on the $E_\nu>\unit[1.6]{GeV}$ flux of upward-going muons from point-source neutrinos as a function of declination. Error bars (too small to see) reflect statistical uncertainty created by averaging over ra. \label{fig:upmufluxlimits}}
\end{figure}

\begin{figure}
  \leavevmode
  \psfig{file=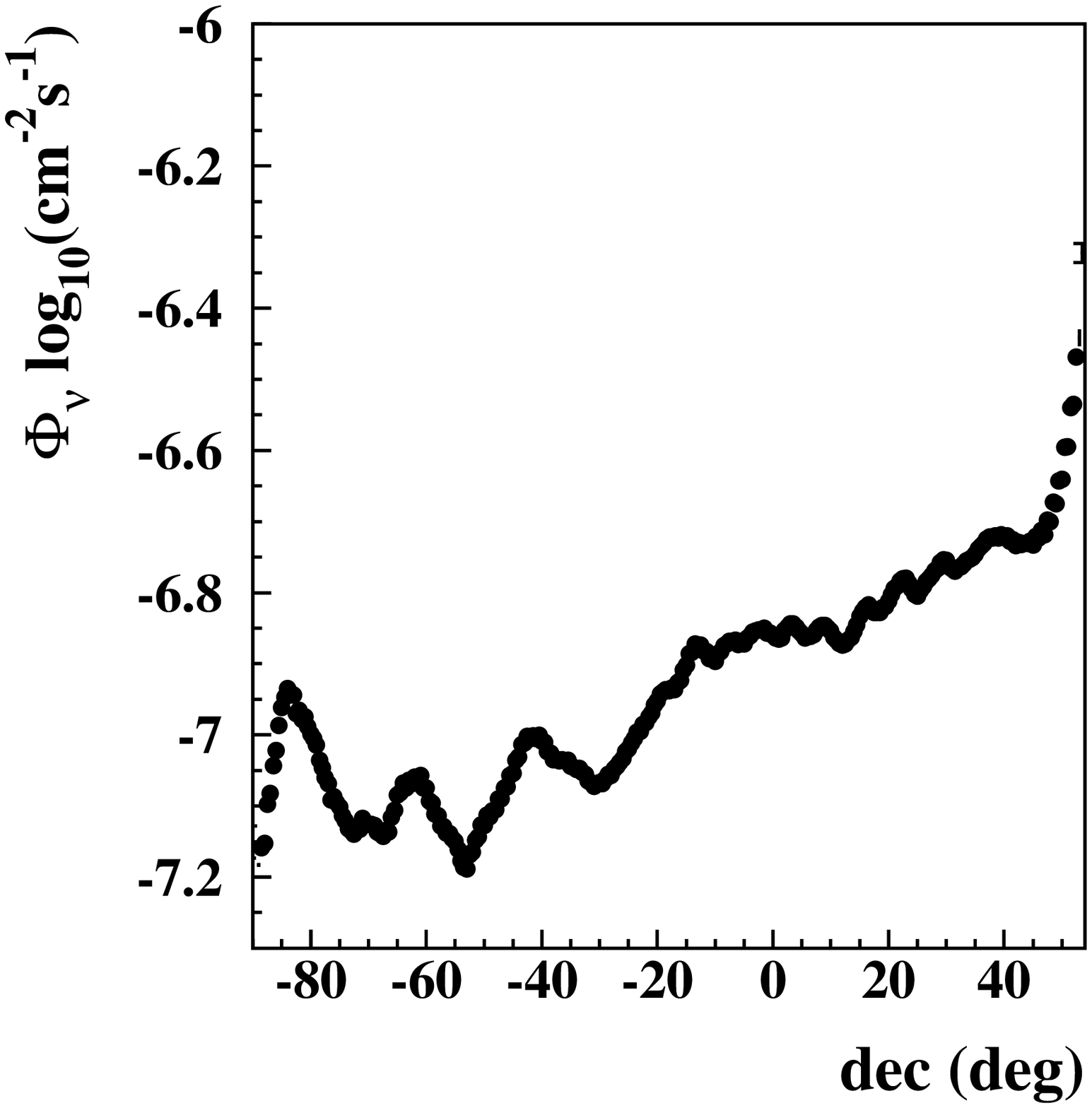,height=2.5in,width=3.0in}
  \caption{90\% CL limits on the $E_\nu>\unit[1.6]{GeV}$ flux of point-source neutrinos as a function of declination. Error bars (too small to see) reflect statistical uncertainty created by averaging over ra.\label{fig:nufluxlimits}}
\end{figure}

\begin{figure}
  \leavevmode
  \psfig{file=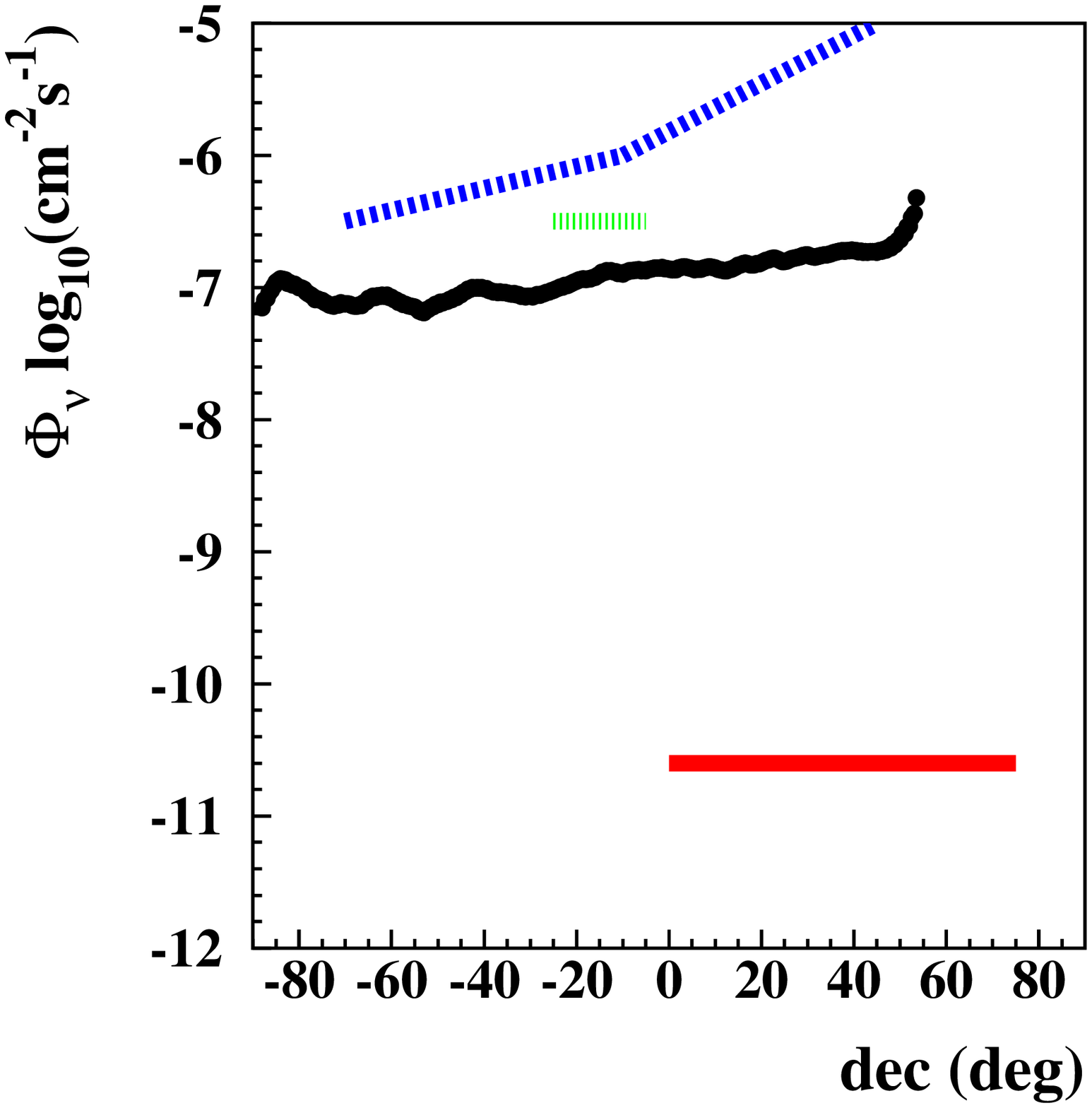,height=2.5in,width=3.0in}
  \caption{A comparison of the neutrino flux limits (at 90\% CL for $E_\nu>\unit[1.6]{GeV}$) obtained here (black data points) with those obtained by AMANDA~\citep{AMANDALimits} for $E_\nu>\unit[1.9]{TeV}$ (approximated by the dotted red line), and limits by MACRO~\citep{MACRO} for $E_\nu>\sim\unit[1]{GeV}$ (approximated by the dashed blue). The dotted green line represents the approximate flux required to produce a signal at $>90\%$ CL with a 50\% detection efficiency. \label{fig:amanda_compare}}
\end{figure}

\begin{figure*}
  \leavevmode
  \psfig{file=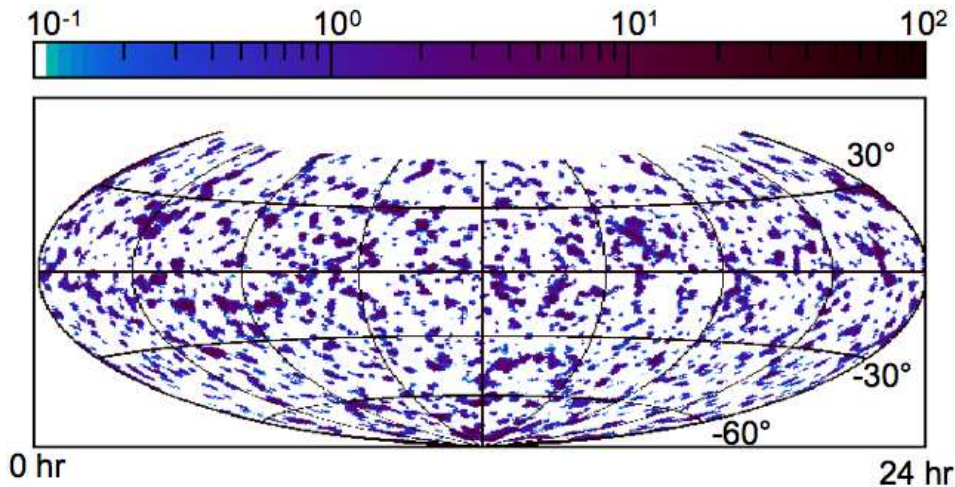,width=5.0in}
  \caption{A sky map of $\Lambda$ in equatorial coordinates.  The maximum value ($\Lambda_\text{max}=19.1$) is below the detection threshold. \label{fig:LambdaMapData}}
\end{figure*}

\begin{figure*}
  \leavevmode
  \psfig{file=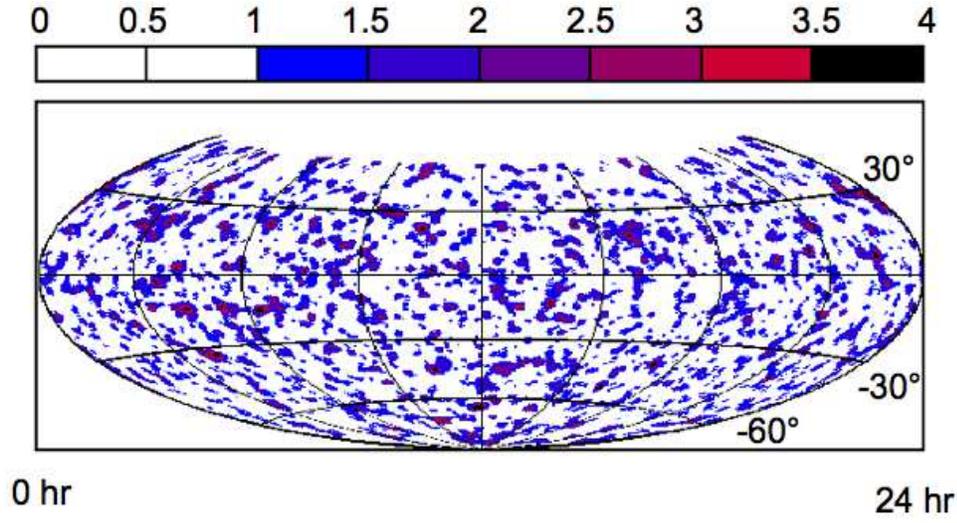,width=5.0in}
  \caption{A sky map of $p(\Lambda)$ in sigma displayed in equatorial coordinates.  \label{fig:probMap}}
\end{figure*}

\begin{figure*}
  \leavevmode
  \psfig{file=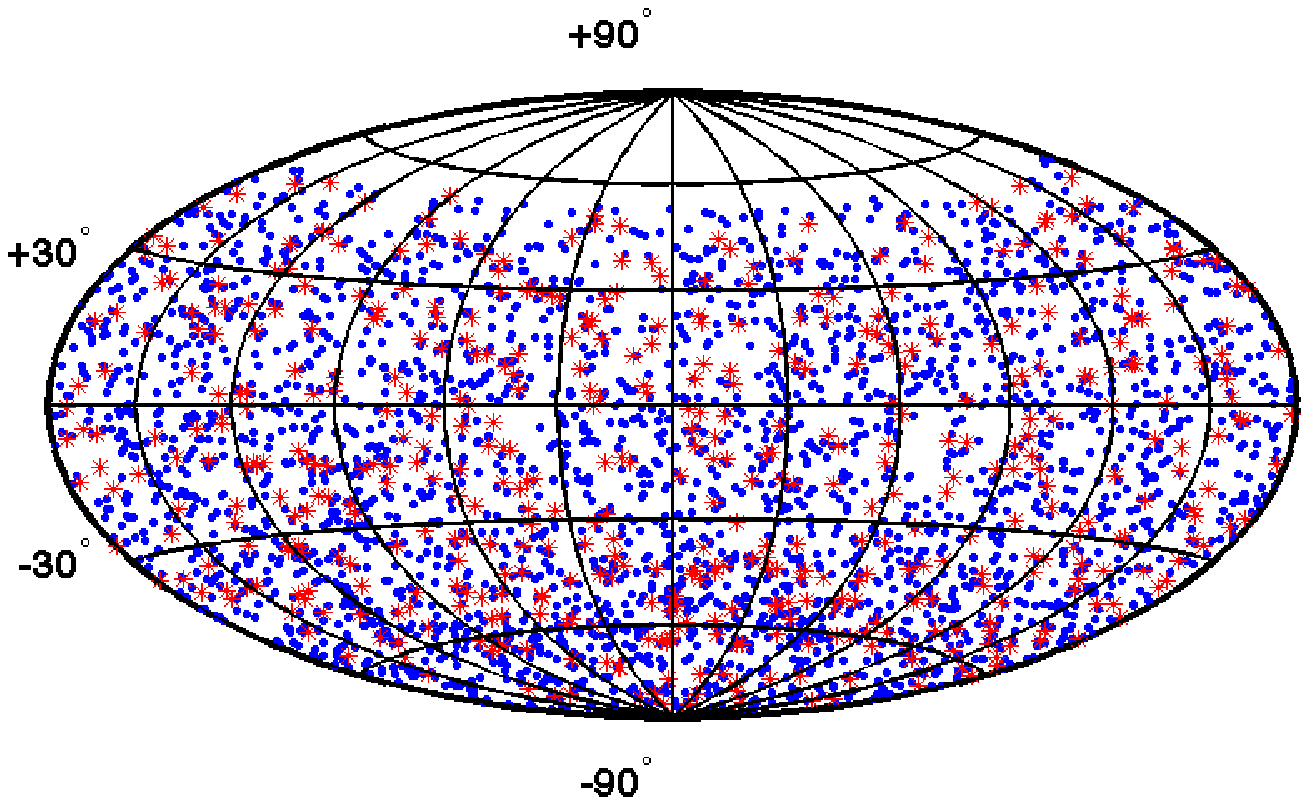,width=5.0in}
  \caption{A sky map in equatorial coordinates depicting each through-going muon event as a dot. Red asterisks are showering muons, blue dots are non-showering. \label{fig:dotmap}}
\end{figure*}

Upper limits on the flux of point source upmus/neutrinos as a function of declination are presented in Figures~\ref{fig:upmufluxlimits} and~\ref{fig:nufluxlimits} respectively.
The limits worsen at higher values of declination due to reduced exposure time.
In Figure~\ref{fig:amanda_compare} we plot the neutrino flux limits obtained in this study (black) alongside an estimate of those obtained by AMANDA~\cite{AMANDALimits} (red) and MACRO~\cite{MACRO} (dotted blue).
Also included in this plot is the expected detection threshold for this study (dotted green) for the case of a 50\% detection efficiency.
We provide a sky map of the likelihood ratio $\Lambda$ in Figure~\ref{fig:LambdaMapData}, in Figure~\ref{fig:probMap} we provide a sky map of $p(\Lambda)$ measured in sigma, and in Figure~\ref{fig:dotmap} we present the through-going muon dataset in the form of a sky map.

\subsection{Suspected Candidates}
\begin{figure}
  \leavevmode
  \psfig{file=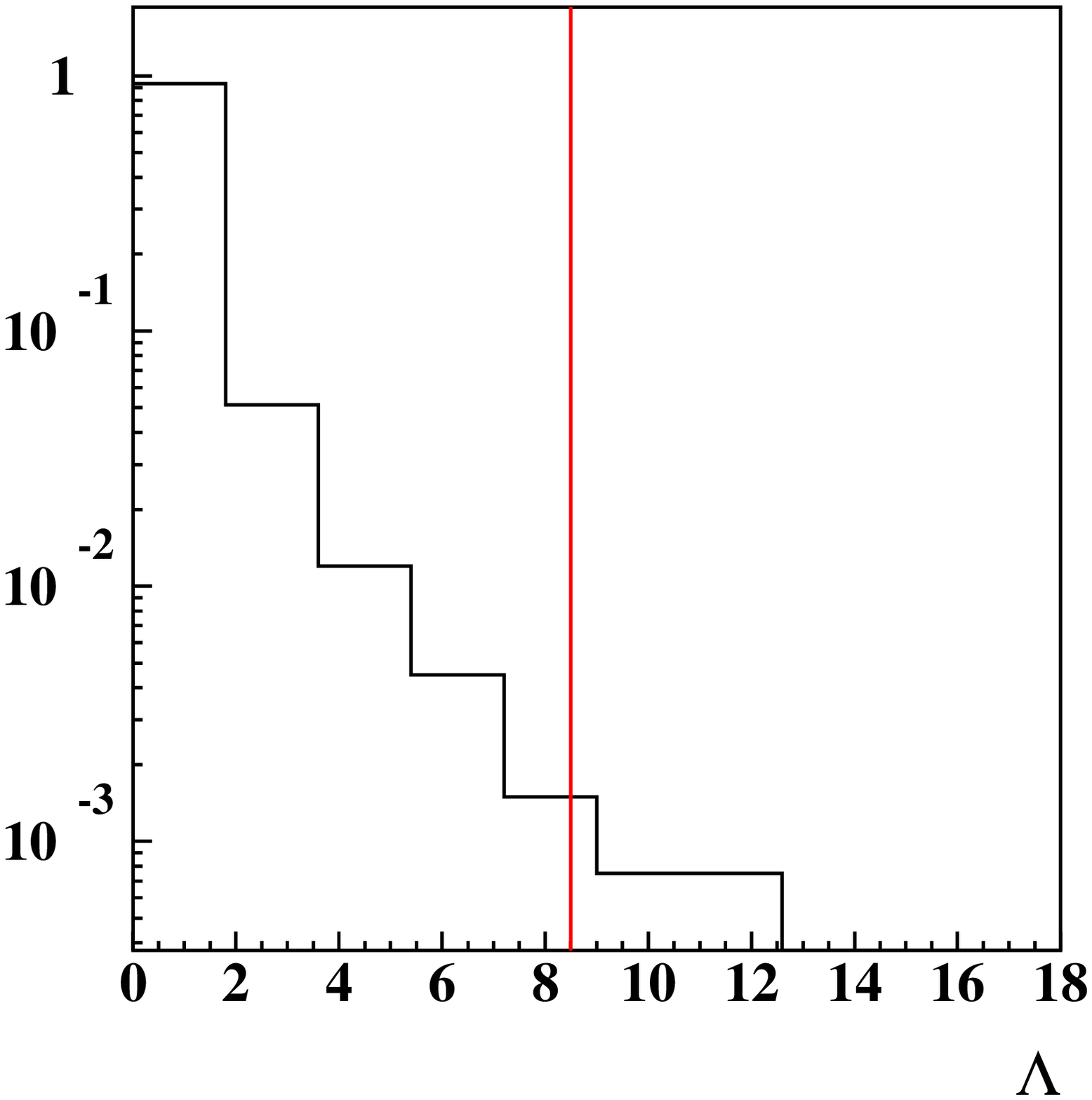,width=2.5in,height=2.5in}
  \caption{A normalized histogram of $\Lambda$ generated for atmospheric (background) events falling inside a search cone centered on RX~J1713.7-3946. The coincident event was measured to have a likelihood ratio of $\Lambda=8.5$ (marked with a red line), which has a 0.2\% chance of being background. \label{fig:snr_lambda}}
\end{figure}

Of the 16 candidates listed in Table~\ref{tab:candidates}, we found a statistically interesting signature from one, the SNR RX~J1713.7-3946, which yielded a signal of $\Lambda=8.5$.
Of the 26 events in the search cone centered on SNR RX~J1713.7-3946, approximately 5 are attributable to signal.
RX~J1713.7-3946 is a one of the few known TeV shell-type SNR.
It is approximately $\unit[1]{kpc}$ away \citep{koyama,fukui} and approximately $\unit[1600]{years}$ old~\citep{wang}.
Using MC, we find the probability of an accidental signal of this magnitude to be 0.16\%, (see Figure~\ref{fig:snr_lambda}.)
Taking into account the fact that we performed sixteen tests, the probability of a chance occurrence becomes 2.5\%.
Limits on the neutrino flux from each candidate (assuming a $\gamma=2$ power law) are recorded in Table~\ref{tab:candidates}.

\begin{deluxetable}{l|l|l|c|r}
  \tabletypesize{\scriptsize}
  \tablecaption{A list of 16 candidates considered to be plausible neutrino point sources with associated limits (at 90\% CL) on $E_\nu>\unit[1.6]{GeV}$ neutrino flux. The abbreviation ``mag'' is for ``magnetar,'' ``pler'' is for ``plerion,'' and ``MQ'' is for ``microquasar.'' Estimating the distance to these sources can be difficult and inexact as distance estimates often depend on measurements of what are assumed to be neighboring objects. The objects in this table are associated with estimated distances that range from $\unit[0.3-50]{kpc}$. \label{tab:candidates}}
 \tablewidth{0pt}
  \tablehead{
    \colhead{source} & \colhead{type} & \colhead{(ra,dec)} &\colhead{$\Phi_\nu^{90\%}$ ($\unit{cm^{-2}s^{-1}}$)}
  }
\startdata
    SGR 1900+14      & mag & ($286.8^\circ,+9.3^\circ$) & $1.12\pm0.12\times10^{-7}$\\
    SGR 0526-66      & mag & ($81.5^\circ,-66.0^\circ$) & $1.15\pm0.13\times10^{-7}$\\
    1E 1048.1-5937   & mag & ($162.5^\circ,-59.9^\circ$) & $6.71\pm0.74\times10^{-8}$\\
    SGR 1806-20      & mag & ($272.2^\circ,-20.4^\circ$) & $1.67\pm0.18\times10^{-7}$\\
    Crab         & pler  & ($83.6^\circ,+22.0^\circ$) & $1.66\pm0.18\times10^{-7}$\\
    Vela X       & pler  & ($128.5^\circ,-45.8^\circ$) & $6.87\pm0.76\times10^{-8}$\\
    G343.1-2.3   & pler  & ($257.0^\circ,-44.3^\circ$) & $6.81\pm0.75\times10^{-8}$\\
    MSH15-52     & pler  & ($228.5^\circ,-59.1^\circ$) & $1.12\pm0.12\times10^{-7}$\\
    RX~J1713.7-3946  & SNR      & ($258.4^\circ,-39.8^\circ$) & $2.67\pm0.29\times10^{-7}$\\
    Vela Jr.         & SNR      & ($133.2^\circ,-46.3^\circ$) & $9.16\pm1.0\times10^{-8}$\\
    MGRO~J2019+37    & SNR      & ($305.2^\circ,+36.8^\circ$) & $2.46\pm0.27\times10^{-7}$\\
    SS433            & MQ & ($288.0^\circ,+5.0^\circ)$ & $1.16\pm0.13\times10^{-7}$\\
    GX339-4          & MQ & ($255.7^\circ,-48.8^\circ$) & $5.50\pm0.61\times10^{-8}$\\
    Cygnus X-3       & MQ & ($308.1^\circ,+40.8^\circ$) & $1.32\pm0.15\times10^{-7}$\\
    GRO J1655-40     & MQ & ($253.5^\circ,-39.8^\circ$) & $1.26\pm0.14\times10^{-7}$\\
    XTE J1118+480    & MQ & ($169.5^\circ,+48.0^\circ$) & $1.29\pm0.14\times10^{-7}$ \\
    \enddata
\end{deluxetable}

\subsection{Correlation with UHE Cosmic Rays}
We found the summed signal from the directions of 27 UHE cosmic rays possibly linked to AGN by the Auger experiment to be $\Lambda_\Sigma=10.3$, which does not constitute a statistically significant signal.
We therefore set a limit on the $E_\nu>\unit[1.6]{GeV}$ flux of neutrinos from Auger UHE directions by averaging the limits obtained for each direction, and found: $\Phi^{90\%}_\nu=\unit[1.06\pm0.12\times10^{-7}]{cm^{-2}s^{-1}}$.

\subsection{Correlation with Gamma-Ray Bursts}
Of the 2200 cataloged GRBs considered, 971 met the criteria of occurring during Super-Kamiokande live time with a zenith angle no greater than $8^\circ$ above the horizon.
The data for this search does not include the recent so-called ``naked-eye'' GRB~080319B, which was the brightest GRB observed to date.
Due to its unusual intensity, a separate search was performed for neutrinos from GRB~080318B~\citep{thrane080319B}, see also~\citep{amanda080319B}.

Of the 971 GRBs visible to Super-Kamiokande, we observed one coincident GRB (designated 991004D) with the upmu dataset.
The upmu occurred $\unit[411]{s}$ after the GRB and was separated by $3.4^\circ$.
GRB~991004D was a long GRB with a duration of $T_{90}=\unit[34]{s}$~\citep{frontera}.
The total and peak gamma-ray fluence were not  atypical nor was the spectral hardness.
Key properties of the coincident GRB are listed in Table~\ref{tab:GRB991004D}.

\begin{deluxetable}{lllll}
  \tabletypesize{\scriptsize}
  \tablecaption{Details of GRB~991004D. \label{tab:GRB991004D}}
 \tablewidth{0pt}
  \tablehead{
    \colhead{ID} & \colhead{ra} & \colhead{dec} & \colhead{UT} & \colhead{JST=UT+9} \\
}
  \startdata
    GRB~991004D & $210.75^\circ$ & $-19.53^\circ$ & 13:13:21 & 22:13:21 \\
    \enddata
\end{deluxetable}

There were 3134 upmus considered in this dataset with a live time of $\unit[2.3\times10^8]{s}$, which implies the mean time between upmu events is $\tau=\unit[7.3\times10^4]{s}$.
Thus, the Poisson probability for detecting one or more background events in the $t\equiv\unit[2000]{s}$ window is given by:
\begin{equation}
  P_\text{time} = 1-P(n=0) = 1-e^{-t/\tau}=0.0270
\end{equation}

The search used an $8^\circ$ half-angle search cone.
Taking into account the fact that the search cone sometimes overlaps with the insensitive region ($z<0$), the effective search cone angle is $6.7^\circ$.
The approximate probability of a coincident background event falling inside the cone is given by:
\begin{equation}
  P_\text{space}=\Omega_\text{cone}/\Omega_\text{SK}
  =0.043/11.37=0.00375
\end{equation}
(Here $\Omega_\text{SK}$ is the solid angle of the sky visible to Super-Kamiokande.)
Thus, the probability of observing one or more background events within the $8^\circ$ cone is $P_\text{coincidence}=P_\text{space} \, P_\text{time}=0.0038\cdot0.027=1.03\times10^{-4}$.

\begin{figure}
  \leavevmode
  \psfig{file=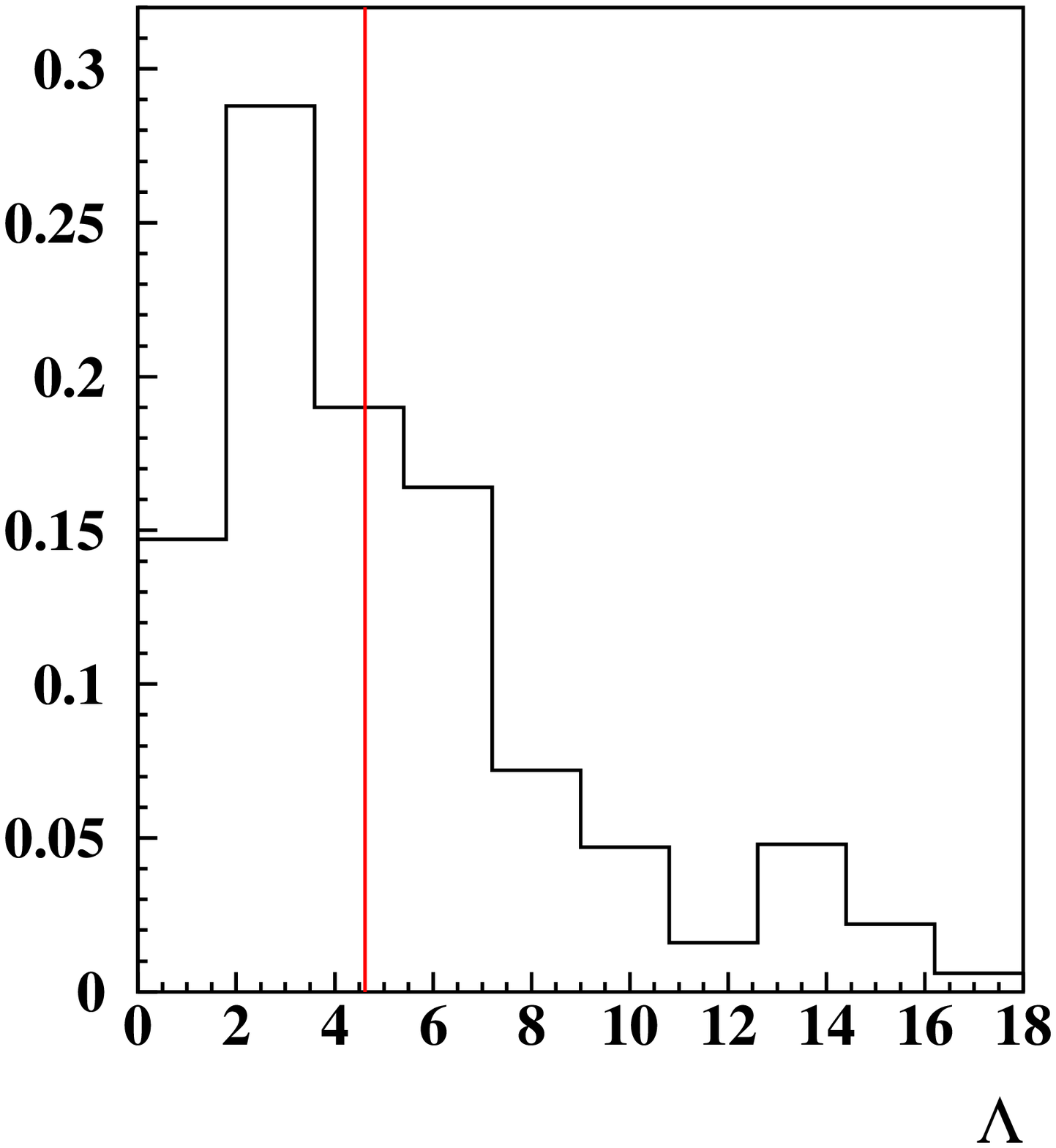,width=2.5in,height=2.5in}
  \caption{A histogram of $\Lambda$ generated for atmospheric (background) events falling inside a search cone centered on a GRB coincidence at (ra,dec)=$(210.75^\circ,-19.04^\circ)$. The coincident event was measured to have a likelihood ratio of $\Lambda=4.6$ (marked with a red line), which has a 48.6\% chance of being background. \label{fig:grb_lambda}}
\end{figure}

The signal strength for the coincident upmu was determined to be $\Lambda=4.6$.
From a MC sample of random background events (see Figure~\ref{fig:grb_lambda}) we determine that the probability of observing $\Lambda\geq4.6$ with background is $P_\Lambda=0.486$.

It follows that the probability for observing a random coincidence with $\Lambda\geq4.6$ in a single trial is $P_\text{trial}=P_\Lambda\, P_\text{coincidence}=5.00\times10^{-5}$.
The probability of observing no coincidence with $\Lambda\geq8.5$ in one trial is, of course, $1-P_\text{trial}=0.99995$.
If we test 971 GRBs, the probability of observing no coincidences with $\Lambda\geq4.6$ in any of the 987 trials is given by $(0.99995)^{971}=0.953$.
Thus, the probability of observing one or more coincidences with $\Lambda\geq4.6$ during 971 trials is $1-0.953=0.047$.
Thus, testing 971 GRBs against the upmu dataset, the single observed GRB-upmu coincidence has a 4.7\% probability of being due to random background.
If we had used a smaller $\pm\unit[500]{s}$ window instead, the confidence level would improve from 95.3\% to 97.6\%.

A previous study by Super-Kamiokande~\citep{SKGRB} found no evidence for a correlation of neutrinos with GRBs and set a $E_\nu>\unit[1.6]{GeV}$ fluence upper limit of $F_\nu^{90\%}>\unit[0.038]{cm^{-2}}$.
Using a $15^\circ$ search cone, Reference~\cite{SKGRB} reported a single coincidence---also GRB~991004D---but was unable to infer a statistically significant signal.
A likely reason for the apparent discrepancy in significance is that the algorithm presented here uses a smaller cone in concert with a likelihood function to filter out events that contribute to the expected background of the $15^\circ$ cone search.

The best limits on GRB fluence currently come from the AMANDA experiment, which reports an upper limit of $\unit[1.4\times10^{-5}]{cm^{-2}}$ for neutrino energies between $\unit[250-10^7]{GeV}$ and assuming a spectral index of $\gamma=2$~\citep{AMANDA}.
In their search, AMANDA employs a timing window defined as the $T_{90}$ start and end times for the burst during which 90\% of the total background-subtracted counts are observed.
(The interval begins when 5\% of the counts have been observed.)
Typically $T_{90}$ is on the order of seconds.
While such a timing window is motivated by the plausible assumptions that GRB neutrinos are emitted at the same times as photons, (and that their journey to Earth is unaltered by any new physics), it is nonetheless true that such a narrow window would not have allowed for detection of the $\unit[411]{s}$ delayed upmu observed in coincidence with GRB~991004D.
The limits calculated by AMANDA, therefore, do not apply to models of GRBs (or new physics) that predict delayed or early neutrino arrival times.

Given AMANDA's limits, along with the fact that there is little in the literature to suggest that GRB~991004D was in any way unusual, we interpret the observed signal as a background fluctuation and thereby obtain a limit on the average $E_\nu>\unit[1.6]{GeV}$ fluence of neutrinos from GRBs to be: $F_\nu^{90\%}=\unit[0.060\pm0.007]{cm^{-2}}$.
This limit is slightly worse than the one obtained in the previous Super-Kamiokande study~\citep{SKGRB} due to the fact that a nearly significant signal necessarily raises the upper limit.
Since the new search algorithm employed here is more sensitive, the apparent signal from GRB~991004D is bigger, and so our limit actually worsens.

\section{Neutrino Flux Limits}\label{nuflux}
In order to compute neutrino flux limits from muon flux limits, we follow \citep{thrane080319B,molly,shantanu} and assume a model for the source spectrum.
It is typically assumed that $d\Phi_\nu/dE\propto E^{-\gamma}$ with $\gamma\approx2$.
Here we assume $\gamma=2$ and investigate the implications of this assumption in Section~\ref{assumptions}.
Given this assumption, the flux of neutrino-induced muons is peaked at high energies, between $\unit[1]{GeV}-\unit[1]{TeV}$ depending on the spectral index (see Figure~\ref{fig:upmuspec}.)

Upmu flux is related to the neutrino flux as in Equation~\ref{eq:nu2mu}.
\begin{equation}\label{eq:nu2mu}
  \Phi_\mu(>E_\mu^\text{min})=\int_{E_\mu^\text{min}}^\infty dE_\nu \,P(E_\nu,\, E_\mu^\text{min}) S(z,\, E_\nu) \, \frac{d\Phi_\nu}{dE_\nu}
\end{equation}
Here $\Phi_\mu(>E_\mu^\text{min})$ is the flux of upmus with energies above the minimum upmu energy of $E_\text{min}\equiv\unit[1.6]{GeV}$.
$P(E_\nu,\, E_\mu^\text{min})$ is the probability that a neutrino with energy $E_\nu$ creates a muon with energy greater than $E_\mu^\text{min}$ and $S(z, E_\nu)$ is the Earth's shadow factor.
We use cross sections from the GRV94 parton distribution function~\citep{Gluck} and the muon range is determined using Reference~\cite{Lipari}~\cite{Reno}.

\begin{equation}\label{eq:nu2muProb}
  P(E_\nu,\, E_\mu^\text{min})=N_A\, \int_0^{E_\nu}dE_\mu\, \frac{d\sigma_{CC}}{dE\mu}(E_\mu,\,E_\nu)\, R(E_\mu,\,E_\mu^\text{min})
\end{equation}
Here $d\sigma_{CC}/dE_\mu(E_\mu,\,E_\nu)$ is the charged current component of the neutrino-nucleon cross section; (neutral current interactions do not produce muons).
$R(E_\mu,\,E_\mu^\text{min})$, meanwhile, is the average range in rock for a muon with an initial energy of $E_\mu$ and a final energy greater than $E_\mu^\text{min}$.

\begin{equation}\label{eq:shadow}
  S(z,\,E_\nu)=e^{-l_\text{col}(z)\,\sigma(E_\nu)\,N_A}
\end{equation}

Here $l_\text{col}(z)$ is the Earth's zenith angle-dependent column depth, and $N_A$ is Avogadro's number scaled by the density of water.
The column depth is calculated using the ``Preliminary Earth Model'' in Reference~\cite{Gandhi96}.

\section{Systematic Error}
Two sources of systematic error are dominant in our calculation of the neutrino flux.
One source is due to uncertainty in the neutrino-nucleon cross section (used in Equation~\ref{eq:nu2mu}), which has an associated uncertainty of $\approx10\%$ at upmu energies \citep{Gandhi96}.
To ascertain how this affects the neutrino flux uncertainty we calculate the flux two ways, $\Phi_+\equiv\Phi(\sigma\rightarrow110\%\,\sigma)$, $\Phi_-\equiv\Phi(\sigma\rightarrow90\%\,\sigma)$, and estimate the uncertainty as half the difference.

Another systematic uncertainty arises from the fact that we only include charged current interactions in our calculation of the Earth shadow.
While neutrinos interacting via charged current interactions produce muons and disappear from the incoming beam, neutrinos interacting via neutral current interactions can lose energy but remain in the beam.
By including only charged current contributions in the Earth shadow, our calculation is too optimistic because neutral current interactions reduce the measured flux at the detector.
If we include both charged and neutral current contributions, however, our calculation is too pessimistic since some neutrinos will reach the detector even after undergoing neutral current interactions.

Therefore, we calculate $\Phi$ two ways: $\Phi_{CC}\equiv\Phi(\sigma=\sigma_{CC})$ and $\Phi_{NC}\equiv\Phi(\sigma=\sigma_{CC}+\sigma_{NC})$.
The neutrino fluence is estimated as the average of $\Phi_{CC}$ and $\Phi_{CC+NC}$ and the uncertainty is estimated as half the difference.
We add these two uncertainties in quadrature and find the total uncertainty in $\Phi_\nu$ to be 11\%.

\section{Assumptions in the Flux Calculation}\label{assumptions}
Before concluding, we assess two assumptions made in the calculation of neutrino flux.
First, we assume a spectral index of $\gamma=2$, and so it is illuminating to address how the flux limits change if we chose a less favorable spectral index.
Repeating the neutrino flux calculations with a more pessimistic spectral index of $\gamma=3$, we find that the flux limits are raised by approximately three orders of magnitude due to the reduced neutrino-nucleon cross section at lower energies.
We note that the two spectral indices yield comparable limits for the flux of $\sim\unit[200]{GeV}$ neutrino-induced muons where the normalized spectra intersect (see Figure~\ref{fig:upmuspec}).
Thus, numerical estimates of neutrino flux are extremely sensitive to assumptions about the source spectrum.
While it is necessary to make an assumption about the nature of the source---and the $\gamma=2$ power law spectrum is certainly a useful standard assumption---it should be noted that until we know more about point-source spectra, numerical limits on neutrino flux are best suited for comparison between experiments.

A second assumption implicit in our flux calculations is that point sources emit neutrinos and not antineutrinos.
The cross section for antineutrino-nucleon interactions is less than the cross section for neutrino-nucleon interactions due to the $V-A$ nature of the weak force, and this causes the limits for antineutrino flux to be 30\% higher than limits obtained for neutrinos.

\section{Conclusions}
We have constructed an algorithm for the purpose of detecting neutrino point sources. 
We find that this algorithm is more sensitive by a factor of two in comparison to a previous point-source search.
Using the algorithm, we performed five tests.
We set limits on $E_\nu>\unit[1.6]{GeV}$ neutrino flux at 90\% CL as low as $\unit[6\times10^{-8}]{cm^{-2}s^{-1}}$ for declinations less than $+54^\circ$.
While not competitive with AMANDA's limits, these limits are among the best for $\text{dec}<0^\circ$.
In a test of suspected sources, we looked for signals from 16 predetermined objects classified as potentially bright neutrino point sources.
Of the 16 objects tested, we found one to have a high enough signal to warrant interest.
The SNR, RX~J1713.7-3946, was found to possess a signature at 97.5\% CL.
Northern hemisphere neutrino observatories such as ANTARES, NEMO, and NESTOR should be able to assess the possibility that the signal is real since the object is located at $\text{dec}=-39.8^\circ$.

Motivated by interest in correlations of AGN with UHE cosmic rays, generated by recent Auger experiment results, we looked for a correlation of upmus with UHE cosmic rays.
We found no evidence for a correlation and set a limit on the average $E_\nu>\unit[1.6]{GeV}$ neutrino flux from Auger UHE event directions: $\Phi_\nu^{90\%}=\unit[1.06\pm0.12\times10^{-7}]{cm^{-2}s^{-1}}$.
We performed a systematic search for correlations of upmus with GRBs in the BATSE, HETE, and Swift catalogs.
We found one coincidence (with GRB~991004D), which constitutes a signal at 95.3\% CL.
We set a limit on average GRB $E_\nu>\unit[1.6]{GeV}$ neutrino fluence, $F_\nu^{90\%}=\unit[0.060\pm0.007]{cm^{-2}}$.

\section{Acknowledgments}
The application of this algorithm to searches for astronomical point sources and many details of its design were motivated by discussions with Professor Thompson Burnett who helped develop a similar algorithm for use with the Fermi-GLAST Gamma-ray Space Telescope, and we gratefully acknowledge his input.
Data on muon range and neutrino cross sections were graciously provided by M. Reno.
  The authors gratefully acknowledge the cooperation of the Kamioka Mining and Smelting Company.
  Super-Kamiokande has been built and operated from funds provided by the Japanese Ministry of Education, Culture, Sports, Science and Technology as well as the U.S. Department of Energy and the U.S. National Science Foundation.
Some participants us have been supported by funds from the Korean Research Foundation (BK21) and the Korea Science and Engineering Foundation.

\bibliographystyle{apj}
\bibliography{ms}

\end{document}